\documentclass[aps, prb, preprint, bibnotes, showpacs, preprintnumbers, amsmath, amssymb, superscriptaddress]{revtex4}

\usepackage{graphicx}
\usepackage{dcolumn}
\usepackage{bm}

\begin{document}

\title{Vertical organic spin valves in perpendicular magnetic fields}

\author{M. Gr\"{u}newald}
\affiliation{Physikalisches Institut (EP3), Universit\"{a}t W\"{u}rzburg, Am Hubland, D-97074
W\"{u}rzburg, Germany} \affiliation{Institute of Physics, Martin-Luther-Universit\"{a}t
Halle-Wittenberg, Von-Danckelmann-Platz 3, D-06120 Halle, Germany}

\author{R. G\"{o}ckeritz}
\affiliation{Institute of Physics, Martin-Luther-Universit\"{a}t Halle-Wittenberg,
Von-Danckelmann-Platz 3, D-06120 Halle, Germany}

\author{N. Homonnay}
\affiliation{Institute of Physics, Martin-Luther-Universit\"{a}t Halle-Wittenberg,
Von-Danckelmann-Platz 3, D-06120 Halle, Germany}

\author{F. W\"{u}rthner}
\affiliation{R\"{o}ntgen Center for Complex Material Systems, Universit\"{a}t W\"{u}rzburg, Am
Hubland, D-97074 W\"{u}rzburg, Germany} \affiliation{Institut f\"{u}r Organische Chemie,
Universit\"{a}t W\"{u}rzburg, Am Hubland, D-97074 W\"{u}rzburg, Germany}

\author{L.W. Molenkamp}
\affiliation{Physikalisches Institut (EP3), Universit\"{a}t W\"{u}rzburg, Am Hubland, D-97074
W\"{u}rzburg, Germany} \affiliation{R\"{o}ntgen Center for Complex Material Systems,
Universit\"{a}t W\"{u}rzburg, Am Hubland, D-97074 W\"{u}rzburg, Germany}

\author{G. Schmidt}
\email[Correspondence to G. Schmidt: ]{georg.schmidt@physik.uni-halle.de}
\affiliation{Physikalisches Institut (EP3), Universit\"{a}t W\"{u}rzburg, Am Hubland, D-97074
W\"{u}rzburg, Germany} \affiliation{Institute of Physics, Martin-Luther-Universit\"{a}t
Halle-Wittenberg, Von-Danckelmann-Platz 3, D-06120 Halle, Germany}

\begin{abstract}
We report the results of magnetoresistance measurements in vertical organic spin valves with the
magnetic field oriented perpendicular to the layer stack. The magnetoresistance measurements were
performed after carefully preparing either parallel or antiparallel in-plane magnetization states
of the magnetic electrodes in order to observe traces of Hanle precession. Due to the low mobility
in organic semiconductors the transit time of spin polarized carriers should allow for precession
of the spins in perpendicular fields which in statistical average would quench the
magnetoresistance. However, in none of the experiments we do observe any change in resistance
while sweeping the perpendicular field, up to the point where the electrode's magnetization starts
to reorient. This absence of Hanle type effects indicates that the magnetoresistance is not based
on the injection of spin polarized electrons into the organic semiconductor but rather on
tunneling through pinholes superimposed with tunneling anisotropic magnetoresistance.
\end{abstract}


\maketitle

\section{INTRODUCTION}

A number of experiments using vertical organic spin valves (OSVs) have been demonstrated over the past
years, for example Refs. [\onlinecite{Dediu2002,Xiong2004,Majumdar2006169,Xu2007,Wang2005,Santos2007,Nguyen2010,Schoonus2009}]. In
many cases it is still unclear whether a tunneling based magnetoresistance effect
(TMR/TAMR)\cite{Santos2007,Schoonus2009,Gruenewald2011} or actual spin injection and consequently
giant magnetoresistance (GMR)\cite{Xiong2004,Nguyen2010} is the origin of the observed effects.
Although the investigation of I/V-characteristics or the temperature dependence of the device
resistance can give indications about the underlying transport mechanisms, final proof is still
missing. Further complications arise from the facts that charge injection into organic
semiconductors is often based on tunneling\cite{Baldo2001} and that intermixing of spin dependent transport and
tunneling anisotropic magnetoresistance\cite{Gruenewald2011} can prevent a clear distinction between TMR and GMR.

In the advent of electrical spin injection into inorganic semiconductors a similar problem existed.
Especially in high mobility semiconductors stray magnetic fields originating from magnetic
contacts can modulate intrinsic magnetoresistance effects and thus mimic the signature of spin
injection. For electrically detected spin injection in inorganic semiconductors the final litmus
test was the investigation of Hanle precession in perpendicular magnetic fields\cite{Lou2007,Huang2007}. It can easily be
shown that the principle of Hanle precession \cite{Johnson1985,Johnson1988} is also applicable for organic spin valves, though
with a slightly different result.

The Hanle effect itself is based on the precession of spins with the Larmor frequency $\omega _L$
induced by a magnetic field $B$ which is oriented perpendicular to  the spin:

\begin{equation}
\omega_{L}=\frac{egB}{2m_e}\label{EqLarmor}
\end{equation}

$m_e$ is the electron rest mass, $g$ the electron's $g$-factor. Thus the spin $\vec{s}$ (phase $\varphi_0$ at time $t=0$) becomes time-dependent and can be described by:

\begin{equation}
\vec{s} = \frac{1}{2}\hbar \left(\begin{array}{c} \cos{\left(\varphi_0 + \omega _L\cdot t\right)} \\ \sin{\left(\varphi_0 + \omega _L\cdot t\right)} \end{array}\right)
\label{EqDelta}
\end{equation}

Usually spin polarized transport in organic or inorganic semiconductors is demonstrated in devices consisting of a non-magnetic transport layer placed between at least two ferromagnetic electrodes in a lateral or vertical arrangement, so called spin valve devices. The resistance $R$ of a spin valve device depends on the relative magnetization of the electrodes (antiparallel $R_{AP}$/parallel $R_P$). Provided that the electrodes exhibit different coercive fields, $R$ therefore can be adjusted by a magnetic field $B_{ip}$ which is applied in the sample plane.\\

If in a spin valve device spins actually are injected into the non-magnetic spacer layer the Hanle
effect can cause a change of the measured signal (device resistance/current). The spins then precess in a magnetic field $B_z$ applied perpendicular to
the sample plane while the in-plane field $B_{ip}$ is kept constant at zero. This precession changes the relative
orientation of the spins with respect to the magnetization of the electrodes. This
change of the spins' direction has the same consequences as a change of the electrodes
magnetization and therefore can be detected in the measured signal.\\

The results that can be obtained in a specific experiment investigating the presence of the Hanle
effect may be varying depending on the transport characteristics of the material under
investigation. However the Hanle effect always appears as spin-dephasing if rather large magnetic
fields $B_z$ are applied. When $B_z$, which causes the spin precession, is increased $\omega _L$
also gets larger, i.e. the precession gets faster. If $\omega _L$ is sufficiently high the single
spins perform several full precessions during their transit through the layer under investigation.
As the time $t_{trans}$ which is needed for the transit through this layer cannot be the same for
every single spin in the material the spin-polarization of the total current averages out due to the
precession. This loss of spin-polarization usually is called
spin-dephasing\cite{Lou2007,Tombros2007}. In materials in which the spin transport occurs rather
incoherently, i.e. the variation in $t_{trans}$ is large, spin-dephasing can already be observed
at small $B_z$ resulting in a simple quenching of the magnetoresistance. In contrast, if coherent
spin transport is present large $B_z$ must be applied in order to observe the decrease in
magnetoresistance while the coherent precession is observed as so called Hanle
oscillations\cite{Huang2007} of the resistance.\\

\begin{figure}

\includegraphics[width=9cm]{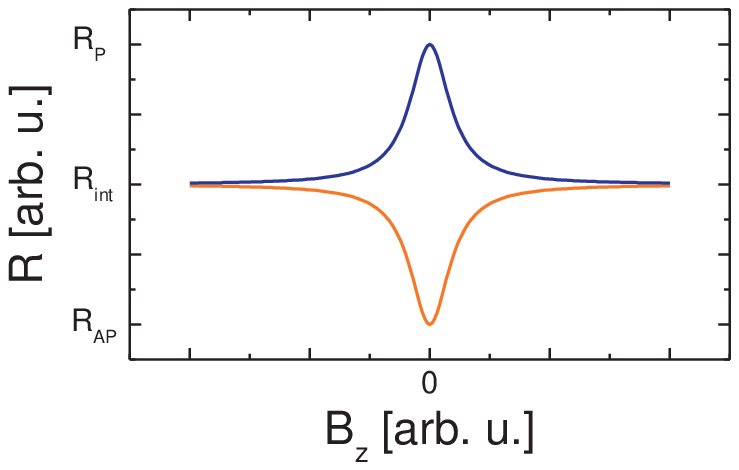}

\caption{Theoretical traces of a device's resistance during experiments in a perpendicular magnetic
field $B_z$ for the previously prepared parallel (blue, dark grey curve) and antiparallel (orange, light
grey curve) spin valve configuration assuming incoherent transport. As with increasing $B_z$ the
spin-polarization is decreased due to spin-dephasing the spin valve signal, i.e. the difference of
the two curves, is quenched at high $B_z$. In both measurements, for the antiparallel and
parallel state, the same intermediate resistance state $R_{int}$ is reached at high $B_z$ starting
at the respective value $R_{AP,P}$ at $B_z=0$. The data is shown for a spin valve device with a
negative effect ($R_{AP}<R_P$) as this behavior is usually observed in organic spin valve
devices.\label{FigHanleEx}}
\end{figure}

Fig. \ref{FigHanleEx} shows resistance traces that can be expected for measurements in
perpendicular magnetic fields in a spin valve device in which the Hanle effect is present and
incoherent transport is dominating. To initialize this kind of measurement, it is necessary to
prepare the parallel (antiparallel) state resulting in a device resistance of $R_P$ ($R_{AP}$) at
$B_{ip}=0$ before the sweep of $B_z$ is started. For a spin valve for which $R_P > R_{AP}$ as is typical
for organic spin valves increasing $B_z$ results in a decrease of $R_P$ and in an increase of
$R_{AP}$. Both curves end up at the same intermediate resistance $R_{int}$ at sufficiently large
$B_z$.\\

It should be noted here that the observation of the Hanle effect is the most robust proof for
electrical spin injection and detection. Other methods to demonstrate the injection of spin polarized carriers like faraday
rotation in inorganic semiconductors\cite{Kikkawa1999} or muon spin rotation\cite{Drew2009} or two photon photoemission
\cite{Cinchetti2009} in organic semiconductors (OSCs) can be used as a proof for spin injection, however, they cannot check whether an electrically detected spin valve signal originates from spin polarized carrier transport or from side effects.
Only the Hanle effect provides the necessary proof that the signal truly originates from the electrical detection of spin polarized carriers. Thus the investigation of the Hanle effect by magnetoresistance measurements (MR) in perpendicular magnetic fields (also named perpendicular geometry hereafter) is an indispensable tool to interpret and understand the results of any spin valve experiment.\\

So far, most of the experiments in which Hanle effect and electrical spin injection is observed are
reported for lateral spin valve devices with inorganic spacers (Si, GaAs, Graphene). Our experiments comprise magnetoresistance measurements in perpendicular magnetic fields on a number of OSV devices with a layer of an OSC as the spin transporting material sandwiched between two ferromagnetic electrodes where the bottom electrode consists of $La_{0.7}Sr_{0.3}MnO_3$ (LSMO) and the top electrode is a ferromagnetic metal. The vertical arrangement of the devices functional layers as well as the choice of LSMO and Co or Co alloys as electrode materials is common to nearly all OSVs reported so far.\\

In the discussion about the Hanle effect in vertical spin valves two arguments are often brought forward in order to explain the possible absence of spin precession: Firstly $B_z$  might be
shielded by the ferromagnetic electrodes' layers in the vertical geometry thus impeding the spin precession by reducing $B_z$ to zero. Secondly the thickness of the OSC layer is usually small compared to the length of a spin-conducting channel in lateral spin valve devices. As a consequence the spin transfer from one electrode to the other (e.g. represented by the transit time $t_{trans}$) might occur on a timescale much shorter than the precession time of the spins if a perpendicular field in the $<1\,T$ regime is applied.\\

We can, however, rule out these unwanted effects in our devices for two reasons. First the high in-plane shape anisotropy of ferromagnetic thin films resulting from their large volume
magnetization ($\approx600\,emu/cm^3$ for LSMO\cite{Ziese2010}, $\approx1400\,emu/cm^3$ for Co \cite{Nishikawa1993} and $\approx1500\,emu/cm^3$  for CoFe\cite{Burkert2004}) has to be taken into account. Due to this shape anisotropy the magnetization remains completely in-plane, even for small perpendicular fields, preventing any shielding of $B_z$.\\

For the transit time the following estimation shows that indeed only rather small $B_z$ are needed
for a sufficiently fast precession of the spins. Assuming $g=2$ for any OSC the precession time
$t_{prec}$ can be calculated from Equ. \ref{EqLarmor}

\begin{equation}
t_{prec}=\frac{2\pi m_e}{eB_z}\label{EqPrec}
\end{equation}

This yields $t_{prec}\approx\,36\,ns$ for $B_z=1\,mT$. $t_{prec}$ has to be compared to the transit time which is needed for a spin to travel through the OSC layer with a thickness $d_{OSC}$. We make a rough estimation for $t_{trans}$ as well:

\begin{equation}
t_{trans}=\frac{d_{OSC}}{v}=\frac{d_{OSC}}{\mu E}=\frac{d_{OSC}^2}{\mu U_{bias}}\label{EqTrans}
\end{equation}

$v$ is the velocity of the spins, $E$ the electric field. Typical values for an OSV experiment are:
mobility $\mu=1\cdot10^{-3}\,cm^2/Vs$, $d_{OSC}=100\,nm$ and applied bias voltage
$U_{bias}=100\,mV$. With these values we obtain $t_{trans}=1\,\mu s\approx30\cdot t_{prec}$ from
Equ. \ref{EqTrans}. Thus even for thin OSC layers the precession at small $B_z$ is sufficiently
fast compared to the transit time to cause a spin-dephasing effect. At the same time the
perpendicular field itself is far too weak to be shielded by the ferromagnetic electrodes. It should be noted that the mobility used for the calculation is an upper estimate and the mobilities that we expect to obtain in the experiment are considerably lower.\\

Furthermore we conclude from the comparison of $t_{prec}$ and $t_{trans}$ that the transport
through the OSC layer is so slow that we do not expect to see any Hanle oscillations but only the
Hanle effect caused by the spin-dephasing (see Fig. \ref{FigHanleEx}). This conclusion is
additionally sustained by the fact that the charge transport in amorphous or polycrystalline OSC
layers is occurring by rather incoherent mechanisms like variable range hopping \cite{Sze2006} or
multiple trapping and release\cite{Horowitz1998}.\\

Although no Hanle effect as such has been observed in organic semiconductors, the demonstration of spin precession of conduction electrons in OSC by paramagnetic resonance \cite{Baker2012} clearly shows that it the underlying physics is the same for organic and inorganic semiconductors.

\section{EXPERIMENTAL DETAILS}

We have fabricated two sets of vertical spin valve structures with different OSC materials as
spacer layer which are also different concerning the electrodes and the fabrication process.\\

The first material under investigation is the \textit{n}-type OSC
\textit{N,N'}-bis(n-heptafluorobutyl)-3,4:9,10-perylene tetracarboxylic diimide [PTCDI-C4F7, Fig.
\ref{FigMR}(b)] which is used due to its excellent properties with respect to charge carrier
mobility and stability in ambient conditions\cite{Oh2007}. PTCDI-C4F7 is a material well suited
for spin valve applications as already has been shown in our previous work\cite{Gruenewald2011}.
The devices based on this material have a bottom contact made of $10-20\,nm$ thick LSMO layers,
grown by pulsed plasma deposition\cite{Bergenti2004} on Strontium Titanate (STO) or Neodymium
Gallate (NGO) substrates, and a CoFe top contact. A series of devices with different OSC layer
thicknesses $d_{OSC}$ ranging from $100\,nm$ to $600\,nm$ was
fabricated.\\

For device fabrication with PTCDI-C4F7 as OSC spacer, first Ti/Au metal stripes are deposited on
the LSMO, using optical lithography, evaporation, and lift-off. These stripes serve as alignment
marks and later as bondpads. A rectangular bottom contact is then patterned into the LSMO layer by
optical lithography and dry etching, leaving the metal contact at one side of the rectangle.
Subsequently, the sample is inserted into the UHV-deposition chamber where a bake-out procedure is
performed at 450 $^\circ$C for 1 hour at an oxygen pressure of 10$^{-5}$ mbar, in order to
compensate under-oxygenation which may occur during the processing. Subsequently, the PTCDI-C4F7
layer and the metal top electrode are deposited under different angles of incidence through a
shadow mask with a rectangular opening. After removing the sample from the UHV chamber, Ti/Au
stripes are deposited through a second shadow mask with striped windows. These metal stripes are
later used as bond pads for the top contacts and also serve as an etch mask for the removal of the
top electrode material between the stripes by dry etching. This approach provides clean,
oxygen-free, and reproducible interfaces which are known to be crucial for working OSV devices.\\

Furthermore we also fabricated devices with the well known and extensively investigated material
Tris(8-hydroxyquinoline)- aluminium(III) [AlQ$_3$, Fig. \ref{FigMR}(a)], a low mobility amorphous
\textit{n}-type OSC. The fabrication process for these devices is different from the one for the
PTCDI-C4F7 devices and will be explained in detail elsewhere. The main differences compared to the
PTCDI-C4F7 devices are the fabrication method of the LSMO electrode (thickness $20\,nm$, grown by
pulsed laser deposition) and the patterning of the devices' active area by means of lithography
instead of shadow masks and dry etching. The thickness of the OSC layer is ranging from $40\,nm$
to $100\,nm$ for the AlQ$_3$ devices.\\

The samples are characterized at various temperatures between $4.2\,K$ and room temperature. Preliminary investigations of the observed MR effects were done in either in a $^4$He flow cryostate or a $^4$He bath cryostate, both equipped with an external room temperature electromagnet ($B_{max}=600\,mT$ respectively $B_{max}=800\,mT$) at various temperatures. The experiments in the perpendicular geometry were conducted at $4.2\,K$ in a $^4$He bath cryostate with a 3D vector magnet in which magnetic fields up to $400\,mT$ can be applied in any direction in space.\\

\section{RESULTS}

\subsection{SPIN VALVE EFFECTS}

\begin{figure}

\includegraphics[width=9cm]{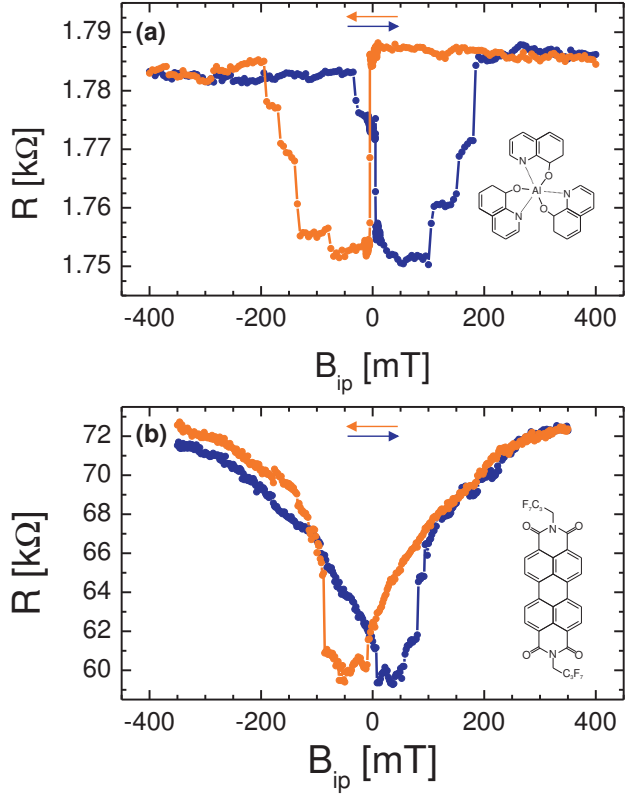}

\caption{Typical MR traces for OSV devices of both types. As indicated by the arrows the sweep from positive saturation to negative saturation is represented by the orange/light grey curve, the opposite sweep direction by the blue/dark grey curve. (a) shows the spin valve behavior of a AlQ$_3$ OSV device ($d_{OSC}=50\,nm$) with a total MR of $\approx-2.0\,\%$. Only sharp switching events between the parallel and antiparallel configuration and back are observed for this device. The measurement is done at $4.2\,K$ and $U_{bias}=-100\,mV$. The MR trace for a PTCDI-C4F7 OSV device ($d_{OSC}=150\,nm$) in (b) with a total MR of $\approx-20\,\%$ taken at $U_{bias}=-10\,mV$ and $4.2\,K$ exhibits spin valve like behavior and a relatively large non-linear background. A constant increase of $R$ with increasing $B_{ip}$ is superimposed to the two distinct switching events between the parallel and antiparallel configuration and back (spin valve signal). The panels also show the molecular structure of the respective OSC material under investigation.
\label{FigMR}}
\end{figure}

Fig. \ref{FigMR} shows typical magnetoresistance traces for both types of devices. The magnetoresistance trace of an AlQ$_3$ based device (present device: $d_{OSC}=50\,nm$) is shown in Fig. \ref{FigMR}(a). The actual spin valve signal is very pronounced and almost no background effect is observed, only a small decrease of $R$ with increasing $B_{ip}$ can be discerned. The spin valve effect itself is negative (low resistance state appears for parallel magnetization of the electrodes $R_{AP}$). The change of the magnetization can be identified as sharp switching events in the MR trace. When $B_{ip}$ is swept from high positive to high negative fields the first switching event leading to the low resistance state $R_{AP}$ appears at small negative fields ($B_{ip}\approx-10\,mT$). This switching can be ascribed to the magnetization reversal of the LSMO electrode. The Co electrode's magnetization is reversed at higher fields ($B_{ip}\approx-100\,-\,-200\,mT$) via multiple steps finally resulting in the high resistance parallel magnetization state $R_P$ again. The MR effect has a magnitude of MR$=(R_{AP}-R_P)/R_{AP}\approx-2.0\,\%$ at $4.2\,K$ and $U_{bias}=-100\,mV$ and is symmetric with respect to $B_{ip}=0\,mT$, i.e. for the opposite sweep direction the same behavior is observed in the positive field range, .\\

The magnetoresistance trace of a PTCDI-C4F7 based device (present device: $d_{OSC}=150\,nm$) has a different shape [Fig. \ref{FigMR}(b)]. The observed effect with a total MR ratio of $\approx-20\,\%$ at $4.2\,K$ and $U_{bias}=-10\,mV$ again is symmetric with respect to $B_{ip}=0\,mT$ but has two distinct components. On one hand the MR trace comprises a non-linear background effect, increasing resistance with increasing $B_{ip}$, which is well known for various OSV devices based on other OSCs reported in the literature\cite{Vinzelberg2008, Xu2007, Majumdar2006, Wang2007} and can be explained by the
magnetic electrodes being saturated at higher magnetic fields. The Organic Magnetoresistance Effect\cite{Francis2004} (OMAR) can be excluded very likely as origin for the background signal in our devices as has been shown in previous studies\cite{Gruenewald2011}. On the other hand we see the actual negative spin valve effect which is less pronounced compared to the AlQ$_3$ OSV due to the superimposed background signal. Comparable to the AlQ$_3$ device the first switching event from high ($R_P$) to low resistance ($R_{AP}$) occurs at small $B_{ip}$. When $B_{ip}$ is further increased the magnetization reversal of the second electrode (CoFe) occurs at $B_{ip}\approx-100\,mT$ which is lower than the magnetic field required to reverse the magnetization of the Co electrode in the AlQ$_3$ devices.\\

The spin valve signals presented in Fig. \ref{FigMR} have been studied extensively with respect to various
parameters (temperature, $U_{bias}$, $d_{OSC}$), which will be presented elsewhere. Nevertheless,
these experiments did not yield reliable proof for the underlying transport mechanisms of the
observed spin valve effects. As already mentioned above more information can be gained by
measurements in perpendicular magnetic fields.\\

\subsection{EXPERIMENTS IN THE PERPENDICULAR GEOMETRY: ALQ$_3$ OSVS}
In order to detect the presence of Hanle precession it is necessary to apply the perpendicular magnetic field to the in-plane antiparallel magnetized and the in-plane parallel magnetized states of the spin valve, respectively.  The fact that OSV often exhibit a negative spin valve effect shows that sign and magnitude of the spin accumulation in the organic semiconductor cannot be related to the magnetization states in a straightforward manner.\\

\begin{figure}

\includegraphics[width=9cm]{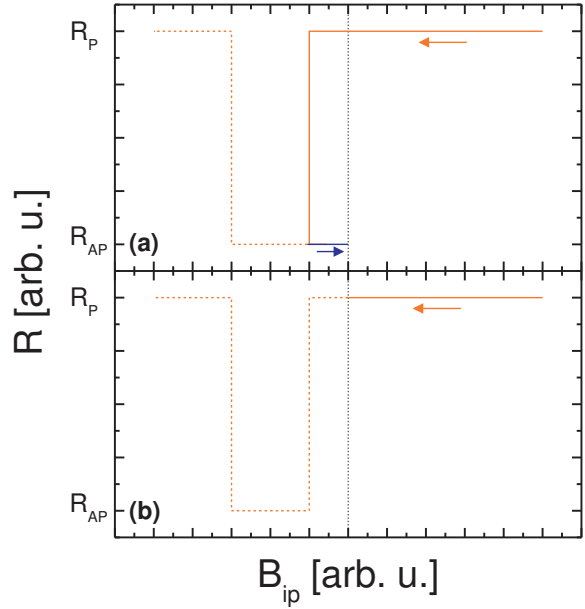}

\caption{Schematic presentation of the $B_{ip}$-sweeps required for the preparation of the antiparallel (a) and parallel (b) magnetization state of a spin valve's electrodes (starting at positive saturation). The solid lines show the sweeps necessary for preparation while the dashed lines represent a complete trace of the negative spin valve signal ($R_{AP}<R_P$). The antiparallel state (a) is initialized by sweeping a minor-loop of $B_{ip}$: $B_{ip}$ is decreased from positive saturation to small negative values until the first switching event occurs (orange/light grey line). Subsequently $B_{ip}$ is returned to 0 (blue/dark grey line) leaving the spin valve in the remanent antiparallel state ($R_{AP}$). The parallel configuration (b) is prepared similarly by sweeping $B_{ip}$ from saturation field to 0. As no reversal of the electrodes' magnetization occurs at positive $B_{ip}$ for this sweep direction the devices resistance remains in the remanent $R_{P}$-state.\label{FigPrepScheme}}
\end{figure}

As a consequence it is not possible to predict whether the influence of spin precession on the two individual resistance states is equally strong. It is however obvious that the difference in resistance between the two states should diminish when Hanle precession is present. Thus it is important to investigate both remanent states in detail. In a recent publication\cite{Riminucci2013}, data has already been presented for magnetic field sweeps perpendicular to the sample plane, however, starting the sweep from perpendicular saturation. In our case the use of a 3D vector magnet allows us to prepare both remanent states individually by running a field sweep from parallel saturation to the desired state and then reducing the in-plane field to zero. Subsequently the perpendicular field is applied and swept over the desired range. After the perpendicular field sweep is finished the in plane MR loop is completed in order to verify that the magnetization state of the electrodes is unchanged (the corresponding field sweeps for the preparation of the spin valves' antiparallel and parallel state are shown in Fig. \ref{FigPrepScheme}).\\

\begin{figure}

\includegraphics[width=18cm]{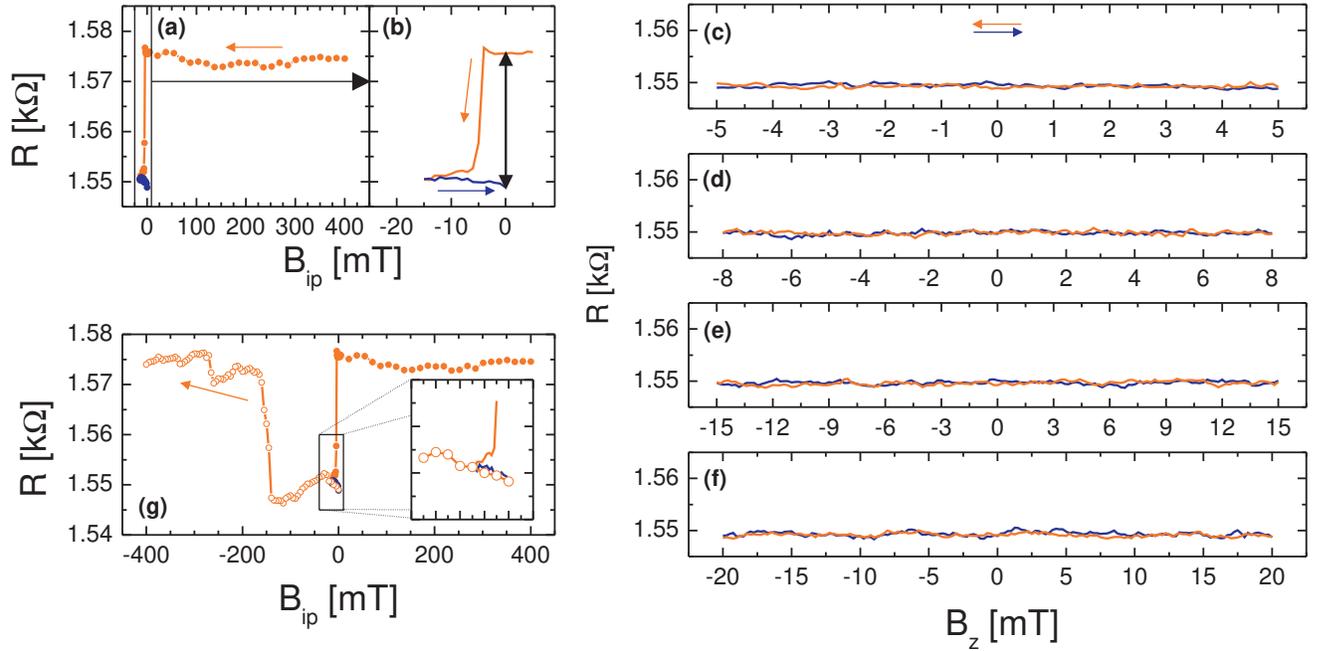}

\caption{ Sequence of measurements investigating the remanent antiparallel state of a AlQ$_3$ OSV in a perpendicular magnetic field, recorded at $4.2\,K$ and $U_{bias}=-100\,mV$. The initial minor-loop shown in (a) and (b) necessary for the preparation of the anti-parallel state, exhibits a MR ratio of $\approx-1.7\,\%$ at $B_{ip}=0\,mT$. (c)-(f) contain the data of the experiments in the perpendicular geometry. Please note that the scale for $R$ is the same in all plots. The maximum absolute $B_z$ and the corresponding minimum precession time $t_{prec}$, calculated using Equ. \ref{EqPrec}, is increased/decreased in every single measurement. We observe a constant device resistance in all measurements (c)-(f) of the series in which the minimum $t_{prec}$ was varied from $\approx7.1\,ns$ to $\approx1.8\,ns$. The statistical variations of $R$ in the range of $\approx\pm0.25\,\%$ are much smaller than the total magnetoresistance effect. Having finished the measurements in the perpendicular geometry the initial minor-loop measurement is completed with a sweep of $B_{ip}$ from $=0\,mT$ to large negative values [line with open circles in (g)]. As this scan clearly reproduces the MR scan in Fig. \ref{FigMR}(a) we can be sure that the magnetization configuration of the electrodes and by this the spin valve state was not disturbed by applying $B_z$.\label{FigHanleALQAP}}
\end{figure}

Fig. \ref{FigHanleALQAP} summarizes a complete sequence of measurements in the perpendicular
geometry for the antiparallel configuration in a AlQ$_3$ based device. All measurements are
performed at $U_{bias}=-100\,mV$ and $T=4.2\,K$. The antiparallel state is prepared by sweeping
$B_{ip}$ on a minor-loop [Fig. \ref{FigHanleALQAP}(a) and (b)]: Starting at high positive fields
$B_{ip}$ is swept beyond $0\,mT$ (orange/light grey curve) until the magnetization of the LSMO
electrode is reversed yielding the low resistance state. Subsequently $B_{ip}$ is set to
$0\,mT$ (blue/dark grey curve). As can be seen in Fig. \ref{FigHanleALQAP}(b) the device remains
in the low resistance state when $B_{ip}$ is turned off. Hence a stable remanent antiparallel configuration
of the electrodes' magnetization with $R_P\approx1.575\,k\Omega$ and $R_{AP}\approx1.55\,k\Omega$
corresponding to minor-loop MR ratio of $\approx-1.7\,\%$ is prepared.\\

The measurements with the magnetic field applied perpendicular to the sample plane involve sweeps
of $B_z$ from $0\,mT$ to higher fields (up to $\pm20\,mT$) and back while the in-plane field
remains at $B_{ip}=0\,mT$. Fig. \ref{FigHanleALQAP}(c)-(f) show the results of these experiments.
The data uses the same scale as the full magnetoresistance sweep.  In each measurement the
maximum absolute value of $B_z$ and by this $\omega_ L$ is increased while the spins' precession
time $t_{prec}$ is decreased. $t_{prec}$ is ranging from
$t_{prec}\approx7.1\,ns$ to $t_{prec}\approx1.8\,ns$ (minimum values
calculated from Equ. \ref{EqPrec} for the respective maximum value of $B_z$). In all measurements we obtain,
independently of the sweep direction of $B_z$ and its maximum value, a constant device resistance
with only statistical variations of $\approx\pm0.25\%$ around $R(0\,mT)$.\\

After returning $B_z$ to zero the minor-loop we have used to prepare the in-plane antiparallel
state is completed. This measurement is done in order to ensure that the electrodes' magnetization
has not been modified by the application of $B_z$ and therefore the antiparallel state of the OSV
device was not disturbed. The line with the open circles in Fig. \ref{FigHanleALQAP}(g) which
shows the data for this measurement clearly reproduces the MR scan for this device recorded
without any interruption [Fig. \ref{FigMR}(a)]. Furthermore this result also allows us to conclude
that the small variations in $R$ observed in the single measurements in Fig. \ref{FigHanleALQAP}(c)-(g)
are not caused by any influence of the applied $B_z$ on the electrodes magnetization.\\

\begin{figure}

\includegraphics[width=9cm]{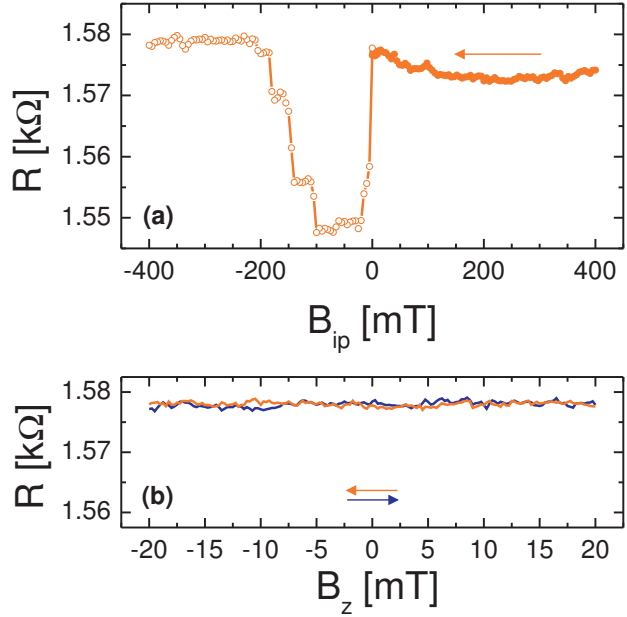}

\caption{Results of measurements in the perpendicular geometry for the remanent parallel state of a AlQ$_3$ OSV recorded at $4.2\,K$ and $U_{bias}=-100\,mV$. A sweep of $B_{ip}$ from high positive fields to $B_{ip}=0\,mT$ is first performed in order to set the device's state of parallel electrodes' magnetization at $B_{ip}=0\,mT$ [closed circles in (a)]. (b) shows data of a measurement in a perpendicular magnetic field for maximum $B_z=20\,mT$ corresponding to a minimum $t_{prec}\approx1.8\,ns$. As for antiparallel configuration the devices resistance is constant apart from statistical variations in the range of $\approx\pm0.25\,\%$. Again the experiment is completed by the termination of the initial MR sweep [open circles in (a)] reproducing the MR trace of the device [Fig. \ref{FigMR}(a)]. \label{FigHanleALQP}}
\end{figure}

The initialization sequence for the parallel configuration is similar to the minor-loop scan in which the antiparallel state is prepared. For the preparation of the parallel state $B_{ip}$ is swept from saturation in the positive field range to $B_{ip}=0\,mT$ ending at $R_P\approx 1.575\,k\Omega$ (Fig. \ref{FigHanleALQP}(a) closed circles).\\

Fig. \ref{FigHanleALQP}(b) shows data of one measurement in the perpendicular geometry with a maximum $B_z$ of $\pm5\,mT$ which is done for the parallel state.
This corresponds to a minimum precession time of $t_{prec}\approx7.1\,ns$. Obviously we do not observe any sizeable change of the OSV's resistance for this configuration as well. After the sweep of $B_z$ the MR loop is completed similarly as for the antiparallel state.
The data of this measurement is shown in  \ref{FigHanleALQP}(a) (open circles) again reproducing
the previously recorded full MR trace.\\

With a minimum precession time of $t_{prec}\approx1.8\,ns$ we are clearly in a limit where the magnetoresistance should be quenched completely by the Hanle effect. The fact that the observed variations in perpendicular fields are statistical and much smaller than the total magnetoresistance clearly indicates that no spin polarized charge transport is responsible for the observed magnetoresistance.

\subsection{EXPERIMENTS IN THE PERPENDICULAR GEOMETRY: PTCDI-C4F7 OSVS}

\begin{figure}

\includegraphics[width=18cm]{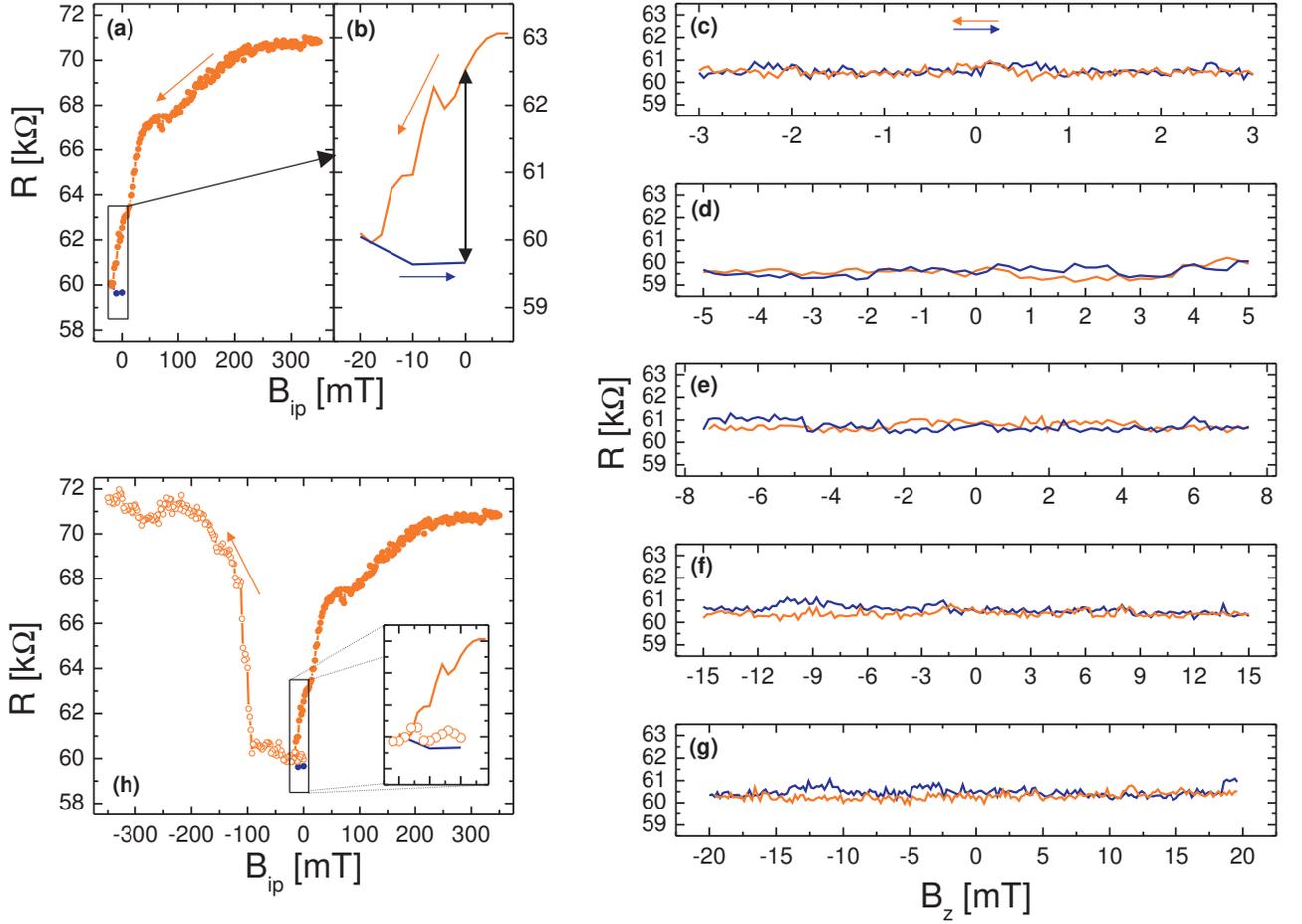}

\caption{Complete measurement sequence of an experiment in the perpendicular geometry for the remanent antiparallel state of
a PTCDI-C4F7 device recorded at $4.2\,K$ and $U_{bias}=-10\,mV$. (a) and (b) show the minor-loop
sweep for the preparation of the antiparallel state with a minor-loop MR ratio of $-4.7\,\%$ at $B_{ip}=0\,mT$. Fig. (c)-(g) comprise the results of the measurements in perpendicular magnetic fields; again the scale for $R$ in (c)-(g) is the same as in (a) and (h). Measurements at $B_{ip}=0\,mT$ with different maximum $B_z$ corresponding to minimum spin precession times from $t_{prec}\approx11.9\,ns$ to $\approx1.8\,ns$ were performed. $t_{prec}$ is calculated using Equ. \ref{EqPrec}. As can be seen in (c)-(g) we only observe statistical variations of the resistance in the range of $\approx\pm1\,\%$. The sequence is finished by completing
the initial minor-loop measurement with a sweep of $B_{ip}$ from $=0\,mT$ to high negative values [line with open circles in (h)]. This measurement shows that the magnetization state of the electrodes was not disturbed by applying $B_z$ as can be seen from a comparison of the data in (h) with the MR scan in Fig. \ref{FigMR}(b).\label{FigHanlePDIAP}}
\end{figure}

For the PTCDI-C4F7 based device similar measurements in perpendicular geometry are performed. Fig.
\ref{FigHanlePDIAP}a shows the preparation of the antiparallel remanent state by reversing the
magnetization of the LSMO layer in a minor loop. As can be seen in the enlarged presentation in
Fig. \ref{FigHanlePDIAP}(b) the minor-loop MR ratio in this device is $\approx-4.7\,\%$ when the
in-plane field is set back to $B_{ip}=0\,mT$. The corresponding resistance values are
$R_{P}\approx62.5\,k\Omega$ for the parallel remanent state and $R_{AP}\approx59.7\,k\Omega$ for the
antiparallel remanent state.\\

The measurements in the perpendicular geometry are shown in Fig. \ref{FigHanlePDIAP}(c)-(g). The
maximum absolute values of $B_z$ correspond to precession times $t_{prec}\approx11.9-1.8\,ns$.
The data in Fig. \ref{FigHanlePDIAP}(c)-(g) shows that for any value of $B_z$ the device
resistance varies by less than $\approx\pm1\,\%$. The sweep completing the initial minor-loop,
is shown in Fig. \ref{FigHanlePDIAP}(h) (line with open circles).\\

\begin{figure}

\includegraphics[width=9cm]{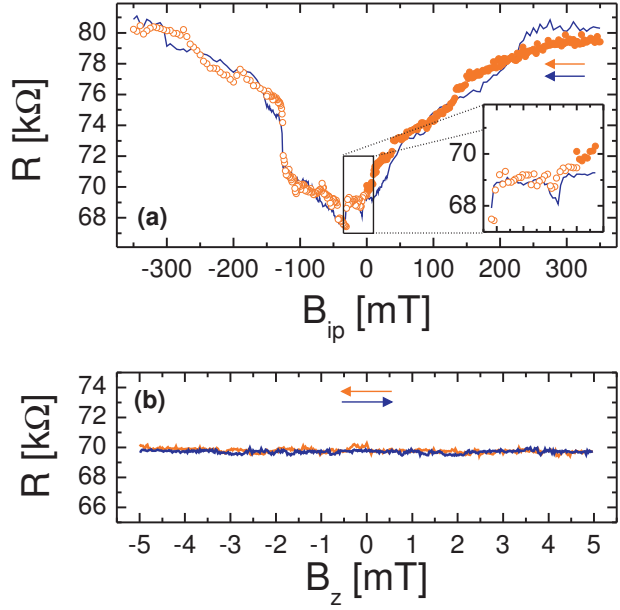}

\caption{Experiments with the magnetic field applied perpendicular to the sample plane for the parallel state of a PTCDI-C4F7 device, recorded at $4.2\,K$ and $U_{bias}=-10\,mV$. The blue (dark grey) line in the background of panel (a) represents the preliminarily recorded full MR sweep. The state of parallel magnetization of the electrodes is prepared by sweeping $B_{ip}$ from high positive fields to $B_{ip}=0\,mT$ [full circles in (a)]. Subsequently the measurements in perpendicular geometry are performed (b). The resistance obtained in a measurement with maximum $B_z=5\,mT$ corresponding to a minimum $t_{prec}\approx7.1\,ns$ is constant and only variations ($\approx\pm1\,\%$) much smaller than the actual spin valve effect are observed. The experiment is
finished by the completion of the initial MR sweep [open circles in (a)] verifying that the magnetization state of the electrodes was not modified by the applied $B_z$. The measured MR trace clearly reproduces the preliminarily recorded curve [blue/dark grey line in (a)].\label{FigHanlePDIP}}
\end{figure}

Fig. \ref{FigHanlePDIP} summarizes the respective experiments in the parallel configuration of the PTCDI-C4F7 device. After sweeping $B_{ip}$ from large positive values to $B_{ip}=0\,mT$ (Fig. \ref{FigHanlePDIP}(a), line with closed circles) measurements with the magnetic field applied perpendicular to the sample plane are performed. Obviously the devices resistance has slightly drifted in the time between the measurements for the antiparallel and the parallel state, however this change did not substantially influence the
devices spin valve performance as was ensured by a full MR sweep (Fig. \ref{FigHanlePDIP}(a) blue/dark grey line in the background of the panel). The maximum $B_z$ for the data in Fig. \ref{FigHanlePDIP}(b) is $\pm5\,mT$ corresponding to a precession time of $t_{prec}\approx7.1\,ns$. This data shows that we do not observe any sizeable change of the device's resistance for the parallel configuration as well. Again the measurement is completed by the final sweep of $B_{ip}$ to negative values. (Fig. \ref{FigHanlePDIP}(a), line with open circles). The data of this measurement continues the initial $B_{ip}$-sweep and reproduces the typical MR trace of the device (Fig. \ref{FigMR}(b), blue/dark grey line in Fig. \ref{FigHanlePDIP}(a) respectively).\\

Although for the PTCDI-C4F7 based device the magnetoresistance trace is less pronounced than for the AlQ$_3$ based one, the variation of the resistance in the parallel and the antiparallel state, respectively, clearly shows that no Hanle precession can be observed, and thus also in this device spin polarized charge transport is not the cause of the magnetoresistance.

\section{CONCLUSION}
We have shown the results of magnetoresistance measurements with the magnetic field applied
perpendicular to the sample plane after preparing in-plane parallel or antiparallel magnetization states.
The devices under investigation differ in OSC material but also in the shape of the observed spin valve signal. In none of our measurements do we observe any Hanle effect in the perpendicular geometry. Although the observations tell nothing about the possible injection of spin polarized carriers into the organic material we must, as a consequence, exclude any spin-polarized transport through the OSC layer as origin of the observed spin valve effects.
Because of the small bias voltages that were used in our experiments we can also exclude fringe-field-induced magnetoresistance that can mimic spin valve signals, however, only at elevated $U_{bias}$\cite{Wang2012}. It is thus likely that TMR by tunneling through pinholes superimposed by TAMR occurring at charge injection into the OSC is the cause of the spin valve behavior of our devices.\\

\section{ACKNOWLEDGEMENTS}
We thank the EU for funding the research in the projects OFSPIN (NMP-CT-2006-033370) and HINTS
(NMP3-SL-2011-263104). We acknowledge Patrizio Grazioso and Alek Dediu from CNR Bologna for
provision of the LSMO layer for the PTCDI-C4F7 based OSV device.

\clearpage

\clearpage

\clearpage

\clearpage

\clearpage

\clearpage

\clearpage

\end{document}